\documentclass[%
reprint,
superscriptaddress,
amsmath,amssymb,amsfonts
aip,
]{revtex4-2}

\usepackage{graphicx}
\usepackage{bm}
\DeclareMathAlphabet{\mathbbold}{U}{bbold}{m}{n}
\usepackage[utf8]{inputenc}

\usepackage{hyperref}
\usepackage[a4paper,margin=20mm]{geometry}
\usepackage{siunitx}
\usepackage{orcidlink}

\date{\today}

\begin{document}

\def\spintec{Univ. Grenoble Alpes, CEA, CNRS, Grenoble-INP, SPINTEC, 38000 Grenoble, France}
\def\dam{CEA, DAM, Le Ripault, F-37260, Monts, France}

\title{Dynamic Landau-Lifshitz-Bloch-Slonczewski equations for spintronics}
\author{Pascal Thibaudeau\,\orcidlink{0000-0002-0374-5038}}
\email{pascal.thibaudeau@cea.fr}
\affiliation{\dam}
\author{Mouad Fattouhi\,\orcidlink{0000-0002-3366-6461}}
\affiliation{\spintec}
\author{Liliana D. Buda-Prejbeanu\,\orcidlink{0000-0002-6105-151X}}
\affiliation{\spintec}

\begin{abstract}
    The atomistic Landau-Lifshitz-Gilbert equation is challenged when modeling spintronic devices where Joule heating is significant, due to its core assumption of a constant magnetization magnitude.
    Based on a statistical framework that treats the magnetization magnitude as a dynamic variable coupled to a thermal bath, we derive a dynamic Landau-Lifshitz-Bloch-Slonczewski set of equations for torques, that captures the transient, heating-induced demagnetization that occurs during high-current operation.
    Integrating these dynamic equations and comparing them to their stochastic equivalents reveals that both the energy landscape and switching dynamics in high-anisotropy systems are similarly modified. This approach yields accurate and accelerated predictions of critical currents and switching times.
\end{abstract}

\maketitle

\section{Introduction}\label{sec:introduction}
The Landau-Lifshitz-Gilbert (LLG) equation, a cornerstone for modeling magnetization dynamics, effectively describes precessional motion and damping influenced by effective magnetic fields and spin-transfer (resp.orbit) torques (STT, resp. SOT)~\cite{ralphSpinTransferTorques2008,baryakhtarLandauLifshitzEquation802015}. However, the LLG equation assumes a constant magnetization magnitude, which limits its applicability in modeling devices subject to significant Joule heating or ultrafast processes where demagnetization effects are pronounced~\cite{rasingUltrafastMagnetizationSwitching2003,hoffmanLandauLifshitzTheoryLongitudinal2013,mondalLaserControlledSpin2018}. This is because temperature significantly impacts the saturation magnetization, and subsequently the anisotropy fields~\cite{callenPresentSatusTemperature1966} and torques, an effect the standard LLG equation neglects.

In spintronic devices like magnetic random-access memories (MRAMs) and spin-torque nano-oscillators (STNOs), high current densities are required for manipulating magnetization~\cite{hirohataReviewSpintronicsPrinciples2020,dienyOpportunitiesChallengesSpintronics2020}. Consequently, substantial Joule heating arises, leading to transient reductions in the magnetization magnitude and altering the energy landscape, switching dynamics, and device performance~\cite{rasingUltrafastMagnetizationSwitching2003,liDynamicsMagnetizationFerromagnet2014,devolderMaterialDevelopmentsDomain2018,hadamekComprehensiveStudyTemperature2023}. Therefore, accurately capturing the interplay between spin-transfer torques and temperature-dependent magnetization is crucial for predicting critical currents, switching times, and overall device reliability~\cite{chaventSteadyStateDynamics2016}.

To address the limitations of the LLG equation under significant thermal excitation, extensions such as the Landau-Lifshitz-Bloch (LLB) formalism have been developed. The LLB formalism incorporates longitudinal relaxation, allowing for dynamic variation of the magnetization magnitude~\cite{garaninFokkerPlanckLandauLifshitzBlochEquations1997,kazantsevaMultiscaleModelingMagnetic2008,atxitiaFundamentalsApplicationsLandau2017}. The LLB equation is an effective framework for analyzing thermal effects in magnetization dynamics. Recent studies have demonstrated its use in developing optimal control strategies for current-induced magnetization dynamics, where adiabatic torque is modulated by the applied current.~\cite{ayouchOptimalControlLandauLifshitzBloch2025}. Nevertheless, a self-consistent derivation of LLB-type equations that incorporates the effects of both temperature and spin-transfer torque is still required to model the magnetization dynamics in realistic spintronic devices~\cite{schiebackTemperatureDependenceCurrentinduced2009,oniciucLandauLifshitzBlochSlonczewskiSimulationsSpintransfertorque2013,lepadatuInteractionMagnetizationHeat2016}.

Building on a statistical mechanics framework that describes how magnetization interacts with a thermal bath~\cite{liThermallyAssistedMagnetization2004,tranchidaHierarchiesLandauLifshitzBlochEquations2018}, we develop dynamic Landau-Lifshitz-Bloch-Slonczewski (dLLBS) equations specifically designed for spintronic systems under strong thermal excitation. This dLLBS framework takes into account both temperature effects and spin-transfer torque, offering a more accurate representation of how magnetization changes over time and better understanding of how high-anisotropy materials switch states. The derived equations account for both longitudinal relaxation due to thermal effects and transverse dynamics induced by the effective field and spin-transfer torque.

\section{Dynamic Landau-Lifshitz-Bloch-Slonczewski equations}\label{sec:derivation}
The dLLBS equations are built upon the stochastic Landau-Lifshitz-Gilbert (sLLG) equation for a single spin, extending it to account for thermal effects and spin-transfer torques consistently, by following statistical protocol.
For a normalized magnetization vector $\bm{s}$, the sLLG equation reads
\begin{align}
    \frac{d\bm{s}}{dt} &= \bm{s} \times \left(\bm{\omega}(\bm{s}) + \bm{h}(t) + \alpha \frac{d\bm{s}}{dt}\right) + \bm{T}_{\text{ST}}(\bm{s})
    \label{sLLG}
\end{align}
where $\bm{\omega}$ is the internal precession frequency, a function of $\bm{s}$ that includes a DC field and a uniaxial anisotropy field with axis $\bm{n}$. It is given by $\bm{\omega}(\bm{s}) \equiv \gamma \bm{B}_{DC} + \gamma B_K(\bm{n}\otimes\bm{n})\bm{s}$, \textcolor{black}{where $\bm{B}_{DC}$ and $B_K$ are the external magnetic field and the anisotropy field amplitude respectively}. Here, $\gamma$ is the gyromagnetic ratio, $\bm{h}(t)$ represents thermal fluctuating field, $\alpha$ is the Gilbert damping parameter, and $\bm{T}_{\text{ST}}(\bm{s})$ is the spin-transfer torque term.
The thermal fluctuation is a stochastic variable and derives from a (vector) Gaussian probability density, such as
\begin{align}
    \begin{aligned}
        \langle\bm{h}(t)\rangle&=\bm{0}\\
        \langle\bm{h}(t)\otimes\bm{h}(t')\rangle&=2D\mathbbold{1}\delta(t-t')
    \end{aligned}
\end{align}
where all the cumulants of higher orders are zero.
Here $D$ is the amplitude of thermal fluctuations that is related to temperature by a (given) thermostat, and $\mathbbold{1}$ is the $3\times 3$ identity matrix. The temperature evolution directly affects the thermal fluctuation amplitude $D$ according to the fluctuation-dissipation theorem, expressed as a Boltzmann weight for classical degrees of freedom.
\begin{align}
    D & = \frac{2\pi\alpha k_B T}{\hbar}
\end{align}
where $k_B$ is the Boltzmann constant. Quantum thermostats can also be considered to better represent the behavior of magnons at low temperatures~\cite{fattouhiUltrafastDynamicsMoments2025}.

When incorporating spin-transfer torques, the most general form of torque is considered. Given a spin polarization direction, $\bm{p}$, the spin torque can be expressed as:
\begin{align}
    \bm{T}_{\text{ST}} &= a_T \bm{s} \times (\bm{s} \times \bm{p}) + b_T \bm{s} \times \bm{p}
    \label{eq:STTtorque}
\end{align}
This expression represents all the possible torques that can be formulated using the basis vectors $\bm{p}$, $\bm{s}$, and $\bm{p} \times \bm{s}$. Here, $a_T$ and $b_T$ are coefficients associated with the damping-like and field-like torques, respectively.
For relevant spintronic applications, these coefficients are current-dependent because the torque itself is generated by the flow of spin-polarized electrons, whose number (and thus the torque strength) scales with the electrical currents~\cite{stilesSpinTransferTorqueDynamics2006}. This is a direct consequence of angular momentum conservation and of the spin transport in magnetic heterostructures~\cite{ralphSpinTransferTorques2008}.

With spin-orbit torques, the momentum $\bm{p}$ is replaced by the spin polarization $\bm{\sigma}$ (the polarization direction of the spin current induced by electron flow in a non-magnet), which is perpendicular to the electron flow. Consequently, the SOT field coefficients $a_T$ and $b_T$ depend on the underlying mechanism, such as the spin Hall effect (SHE), Rashba effect, or inverse spin galvanic effect (iSGE)~\cite{manchonCurrentinducedSpinorbitTorques2019}.

To derive equations governing the ensemble-averaged magnetization, we average Equation~\eqref{sLLG} over the thermal fluctuations, as described in reference~\cite{tranchidaHierarchiesLandauLifshitzBlochEquations2018}.
\textcolor{black}{The key steps are detailed in the Appendix~\ref{appendix:derivation}.}
Introducing $\bm{S}(t) = \langle \bm{s}(t) \rangle$ as the ensemble-averaged magnetization and $\Sigma(t)= \langle \bm{s}(t)\otimes\bm{s}(t) \rangle$ as the correlation matrix, we derive a differential equation on the average magnetization
\begin{align}
    \begin{aligned}
        (1+\alpha^2)\frac{d\bm{S}}{dt}=&\,\bm{S} \times \bm{u} + \alpha \left(\Sigma\bm{u} - \textrm{Tr}(\Sigma)\bm{u}\right) \\
        &-\frac{2D}{(1+\alpha^2)}\bm{S}+\bm{T}_{\text{S}}^{\text{eff}}
    \end{aligned}
    \label{dLLB1}
\end{align}
where $\bm{T}_{\text{S}}^{\text{eff}}$ is the effective spin-transfer torque term, now a function of both $\bm{S}$ and ${\Sigma}$.
Here, $\textrm{Tr}$ stands for the trace and $\bm{u}$ represents the vector in the precession and torque that is independent of $\bm{s}$, expressed as $\bm{u} \equiv \gamma\bm{B}_{DC}+b_T\bm{p}$.
This yields an effective spin torque contribution:
\begin{align}
    \bm{T}_{\text{S}}^{\text{eff}} & = \bm{T}^{(1)}_{\text{S}}(\Sigma)+\bm{T}^{(2)}_{\text{S}}(\Sigma,\bm{S})
\end{align}
where $\bm{T}^{(1)}_{\text{S}}$ and $\bm{T}^{(2)}_{\text{S}}$ are vectors that depend on the correlation matrix $\Sigma$ and the average magnetization $\bm{S}$. To express these vectors explicitly, the linear (both symmetric and antisymmetric) dependence of $\bm{s}$ in the precession field is gathered in a matrix $\mathfrak{M}$. Specifically, $\mathfrak{M} \equiv \gamma B_K \bm{n}\otimes\bm{n} - a_T[\bm{p}]_\times$, incorporating both the anisotropy and spin torque contributions.
We introduce $[\bm{p}]_\times$, a skew-symmetric matrix formed from vector $\bm{p}$ and expressed as
\begin{align}
    [\bm{p}]_\times&\equiv
    \begin{pmatrix}
        0 & -p_3 & p_2 \\
        p_3 & 0 & -p_1 \\
        -p_2 & p_1 & 0
    \end{pmatrix}
\end{align}
This matrix is useful to construct matrix products applying either on vectors or matrices.
For vectors, this is equivalent of a vector product. For instance, we have $[\bm{p}]_\times\bm{S}=\bm{p}\times\bm{S}$.
Thus, we find
\begin{align}
    T^{(1)}_{S,i}\equiv&\,\epsilon_{ijk}(\mathfrak{M}\Sigma)_{kj}\\
    \bm{T}^{(2)}_S\equiv&\,\alpha\left[
        \textrm{Tr}(\Sigma-2\Gamma)(\textrm{Tr}(\mathfrak{M})\bm{S}+\mathfrak{M}\bm{S})\right.\nonumber\\
    &\left.-2((\Sigma \mathfrak{M})^A+[\mathfrak{M},\Sigma])\bm{S}\right]
\end{align}
where $\Gamma\equiv \bm{S}\otimes\bm{S}$, $(\Sigma \mathfrak{M})^A$ is the antisymmetric part of $\Sigma \mathfrak{M}$ matrix, $[\mathfrak{M},\Sigma]$ is the commutator of $\mathfrak{M}$ and $\Sigma$ and $\epsilon_{ijk}$ is the Levi-Civita (or permutation) symbol.

For completeness, the correlation matrix is also tracked, and we find
\begin{align}
    \begin{aligned}
      (1+\alpha^2)\frac{d\Sigma}{dt}&=[[\bm{u}]_\times,\Sigma]-\frac{2D}{(1+\alpha^2)}\left(3\Sigma-\textrm{Tr}(\Sigma)\mathbbold{1}\right)\\
      &-2\alpha\left(
        2[C^A,\Sigma]+\textrm{Tr}(C^S)(\Sigma-2\Gamma)\right.\\
    &\left.-\textrm{Tr}(\Sigma-2\Gamma)C^S\right)
      +T_{\Sigma}^{\text{eff}}
    \end{aligned}
    \label{dLLB2}
\end{align}
The matrix $C^S\equiv\frac{1}{2}\left(C^T+C\right)$ (resp.~$C^A\equiv\frac{1}{2}\left(C^T-C\right)$) is the symmetric (resp.~antisymmetric) part of the tensor product of the field $\bm{u}$ and average magnetization $\bm{S}$, i.e. $C\equiv\bm{u}\otimes\bm{S}$.
We quote $C^T$, the transpose of the $C$ matrix.

The effective tensor $T_{\Sigma}^{\text{eff}}$ is twice the symmetric part of $T_{\Sigma}$, i.e. $T_{\Sigma}^{\text{eff}}\equiv 2T_{\Sigma}^S$, where $T_{\Sigma}$ is given by
\begin{align}
    \begin{aligned}
        T_\Sigma&\equiv\alpha\left(
            2\textrm{Tr}(\Gamma\mathfrak{M})\Gamma-\Sigma(\mathfrak{M}+\mathfrak{M}^T)\Sigma-\textrm{Tr}(\Sigma\mathfrak{M})\Sigma\right.\\
            &\left.-2\textrm{Tr}(\Gamma)\mathfrak{M}\Sigma+2\mathfrak{M}\Sigma\Sigma+\textrm{Tr}(\Sigma)\mathfrak{M}\Sigma\right.\\
            &\left.+{\bm F}\otimes{\bm S}-[\bm S]_{\times}\mathfrak{M}\Sigma+[\mathfrak{M}{\bm{S}}]_{\times}\Sigma\right)
    \end{aligned}
    \label{TSigma}
\end{align}
In Eq.\eqref{TSigma}, ${\bm F}$ is a vector constructed from the components of $\mathfrak{M}$, $\Sigma$, and $\Gamma$ matrices, that are summed as
\begin{align}
    F_i&\equiv\epsilon_{ial}\mathfrak{M}_{ak}(\Sigma_{kl}-2\Gamma_{kl}).
\end{align}
When the matrix $\mathfrak{M}$ is zero (i.e. no anisotropy and damping-like torque), eqs.~\eqref{dLLB1} and \eqref{dLLB2} reduce to those derived in reference~\cite{fattouhiUltrafastDynamicsMoments2025}.

Henceforth, and without loss of generality, we focus on the treatment of spin-transfer torque. What distinguishes these dLLBS equations from conventional approaches is the statistical treatment of thermal fluctuations and their coupling to the spin-transfer torques. Unlike the standard LLB equations that typically assume fixed temperatures, and laws of transformation for material parameters, this formulation allows a dynamic evolution of temperature, without prior knowledge of material parameters.

The equations~\eqref{dLLB1} for $\bm{S}$, and \eqref{dLLB2} for $\Sigma$, constitute our dLLBS formalism, enabling accurate modeling of magnetization dynamics in spintronic devices where thermal effects significantly influence the switching behavior.

\section{sLLG versus dLLBS analysis}

In the sLLG framework, the magnetization magnitude is fixed $(\forall t,\|\bm{s}(t)\| = 1)$, with thermal effects introduced solely via stochastic noise. The reported magnetization is thus an ensemble average over thermal realizations, reflecting changes in both direction and magnitude -- not the instantaneous state.
By contrast, the dLLBS approach permits dynamic evolution of both quantities.
The longitudinal relaxation introduces qualitatively different behavior, particularly near the Curie temperature for a ferromagnet, where $\|\bm{s}\|$ can vary significantly on the timescale of the transverse dynamics.

Thus, to highlight the relevance of the dLLBS approach, we compare the dynamics of the average magnetization vector components $\langle s_x \rangle$, $\langle s_y \rangle$, and $\langle s_z \rangle$ under identical conditions but different thermal treatments, focusing on the simplest cases.

\begin{figure}[htbp]
    \begin{tabular}{c}
        \begingroup
        \resizebox*{0.13\hsize}{!}{$T=1\unit{\kelvin}$}
        \endgroup\\
        \includegraphics[width=0.68\columnwidth,keepaspectratio=true]{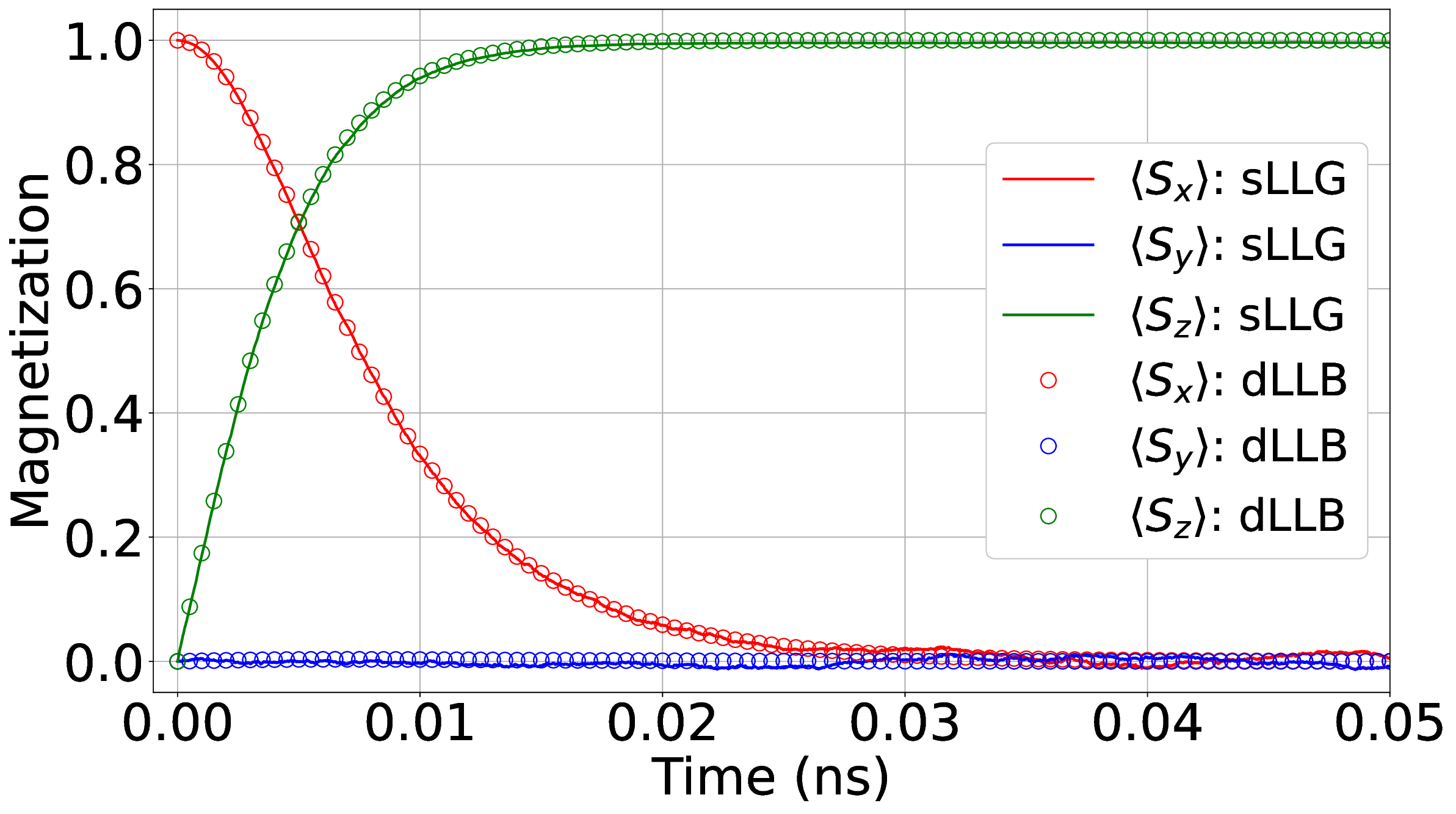}\\
        \begingroup
        \resizebox*{0.15\hsize}{!}{$T=50\unit{\kelvin}$}
        \endgroup\\
        \includegraphics[width=0.68\columnwidth,keepaspectratio=true]{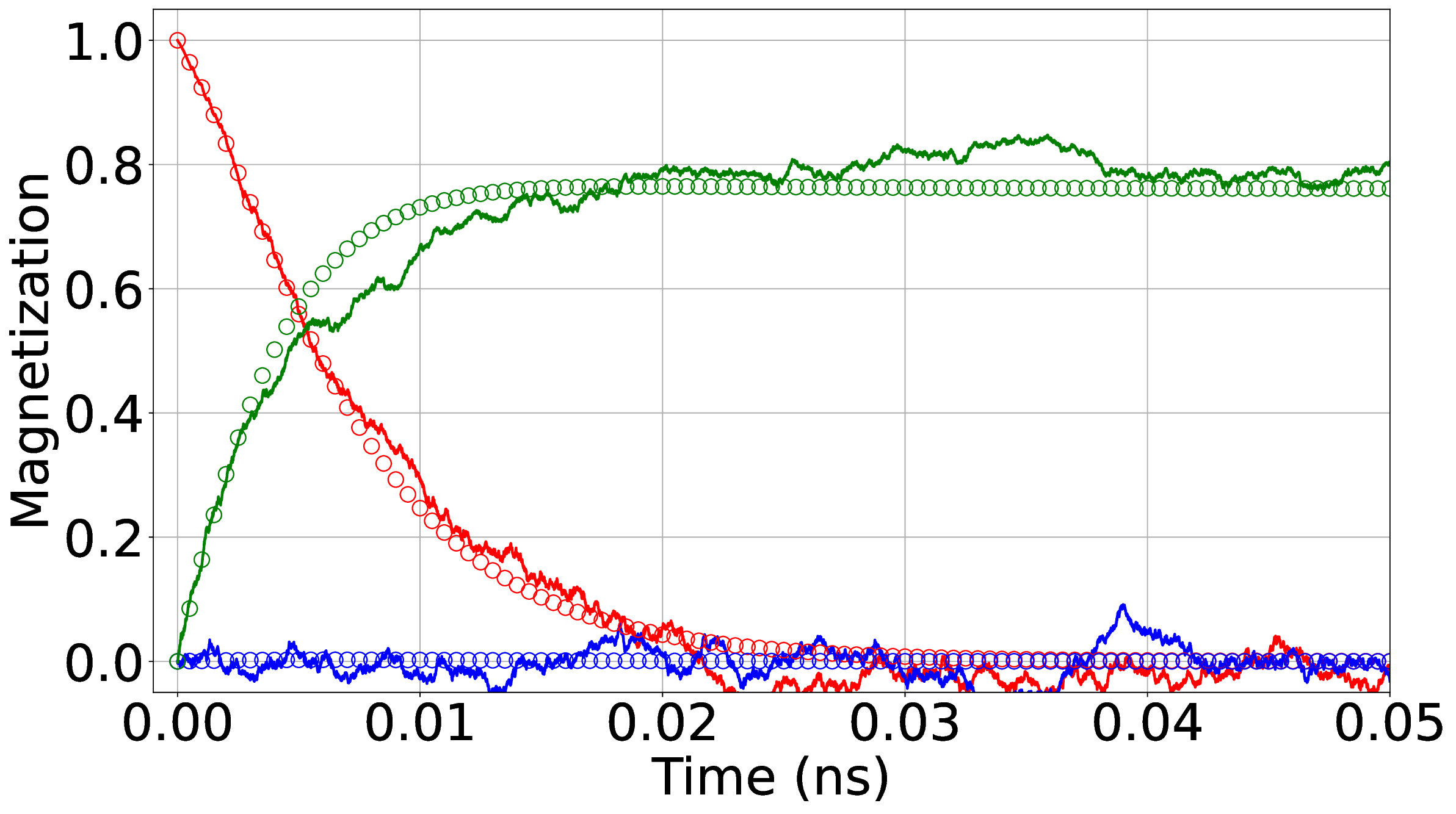}\\
        \begingroup
        \resizebox*{0.15\hsize}{!}{$T=100\unit{\kelvin}$}
        \endgroup\\
        \includegraphics[width=0.68\columnwidth,keepaspectratio=true]{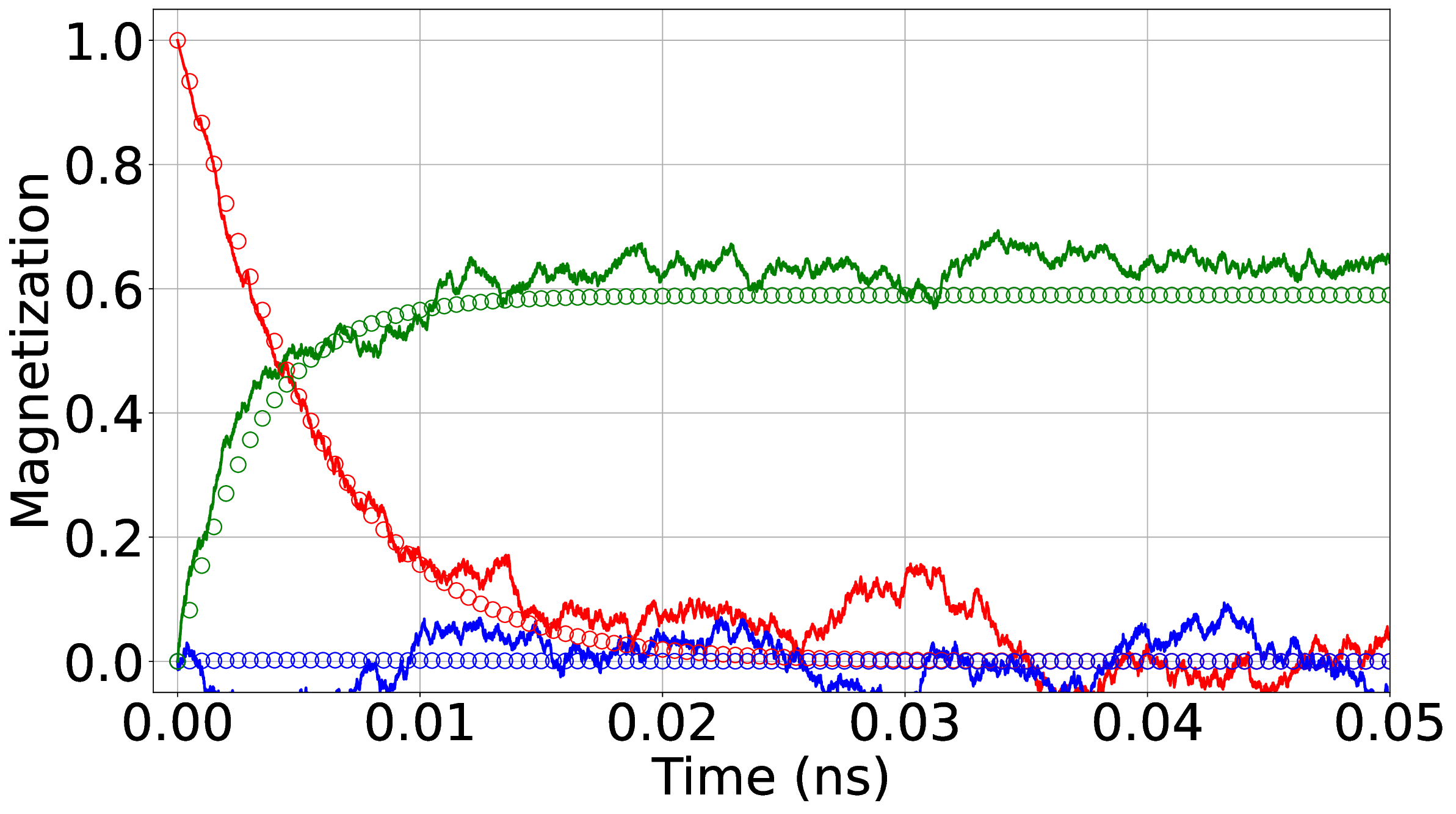}
    \end{tabular}
    \caption{Dynamics of the average magnetization switching of a free MTJ at different temperatures. The average is performed on 100 stochastic realizations of the sLLB equation (Eq.\eqref{sLLG}) (lines) and compared to a single simulation using the dLLBS equations (Eqs.\eqref{dLLB1},~\eqref{dLLB2}) (dots) (see text).}
    \label{fig:sllG_dLLBS}
\end{figure}

First, the free layer of a representative magnetic tunnel junction (MTJ) is modeled as a single macrospin subject to the influence of a fixed polarizer. This is consistent given the small dimensions of the typical MTJs~\cite{vincentAnalyticalMacrospinModeling2015}.
The results, displayed in the figure~\ref{fig:sllG_dLLBS}, are compared by averaging $\sim{\!\! 100}$ stochastic traces obtained from solving the sLLG equation with a single simulation from the dLLBS approach.
This figure highlights the ultrafast dynamics of the average magnetization, initially aligned along the $\hat{\bm{x}}$ axis, under the influence of a damping-like torque with parameters characteristic of typical MTJs~\cite{fengSuperparamagnetismMgObasedMagnetic2010} (i.e $a_T\approx -1\unit{\tesla}$, ${\bm p}=\hat{\bm{z}}$, $\alpha=0.005$, and ${\bm{B}_{DC}=\bm{B}_K=\bm{0}}$).
In these simulations, magnetic exchange between magnetic moments is neglected, which simplifies the interpretation of the resulting macrospin dynamics. Consequently and because it is a single macrospin, the system exhibits a paramagnetic behavior as temperature increases, characterized by magnetization reversal driven by spin-transfer torque and followed by relaxation of the magnetization component to its thermal equilibrium state.
This behavior is accurately reproduced by each dLLBS simulation.
In Figure~\ref{fig:sllG_dLLBS}, the dynamics of the average magnetization at $T=1\unit{\kelvin}$ mirror those in Figure a) of Reference~\cite{meoSpintransferSpinorbitTorques2023}, provided the polarizer and initial magnetization directions are swapped.

\begin{figure}[htbp]
    \begin{tabular}{c}
        \begingroup
        \resizebox*{0.13\hsize}{!}{$T=0\unit{\kelvin}$}
        \endgroup\\
        \includegraphics[width=0.7\columnwidth,keepaspectratio=true]{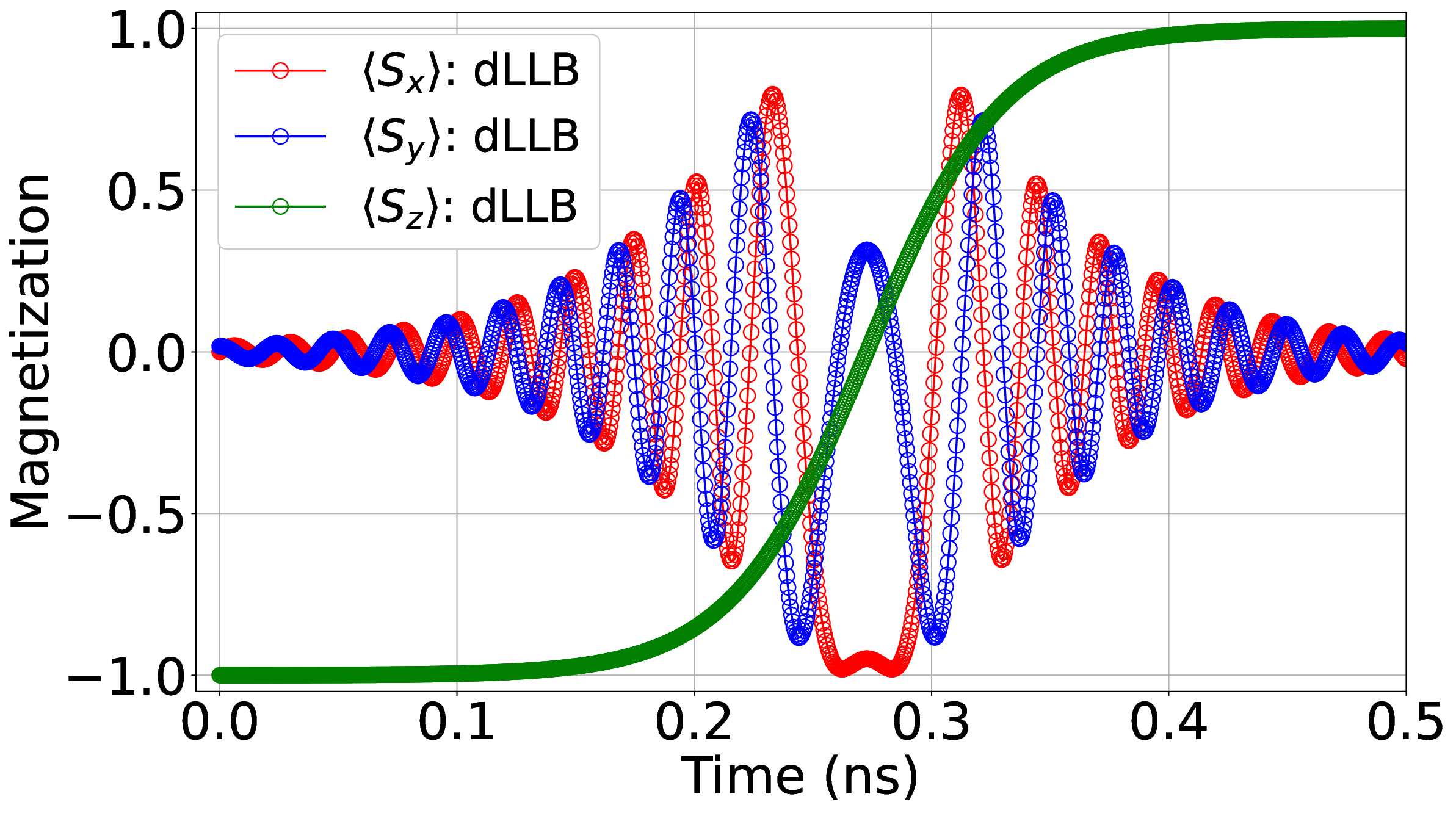}\\
        \begingroup
        \resizebox*{0.15\hsize}{!}{$T=10\unit{\kelvin}$}
        \endgroup\\
        \includegraphics[width=0.7\columnwidth,keepaspectratio=true]{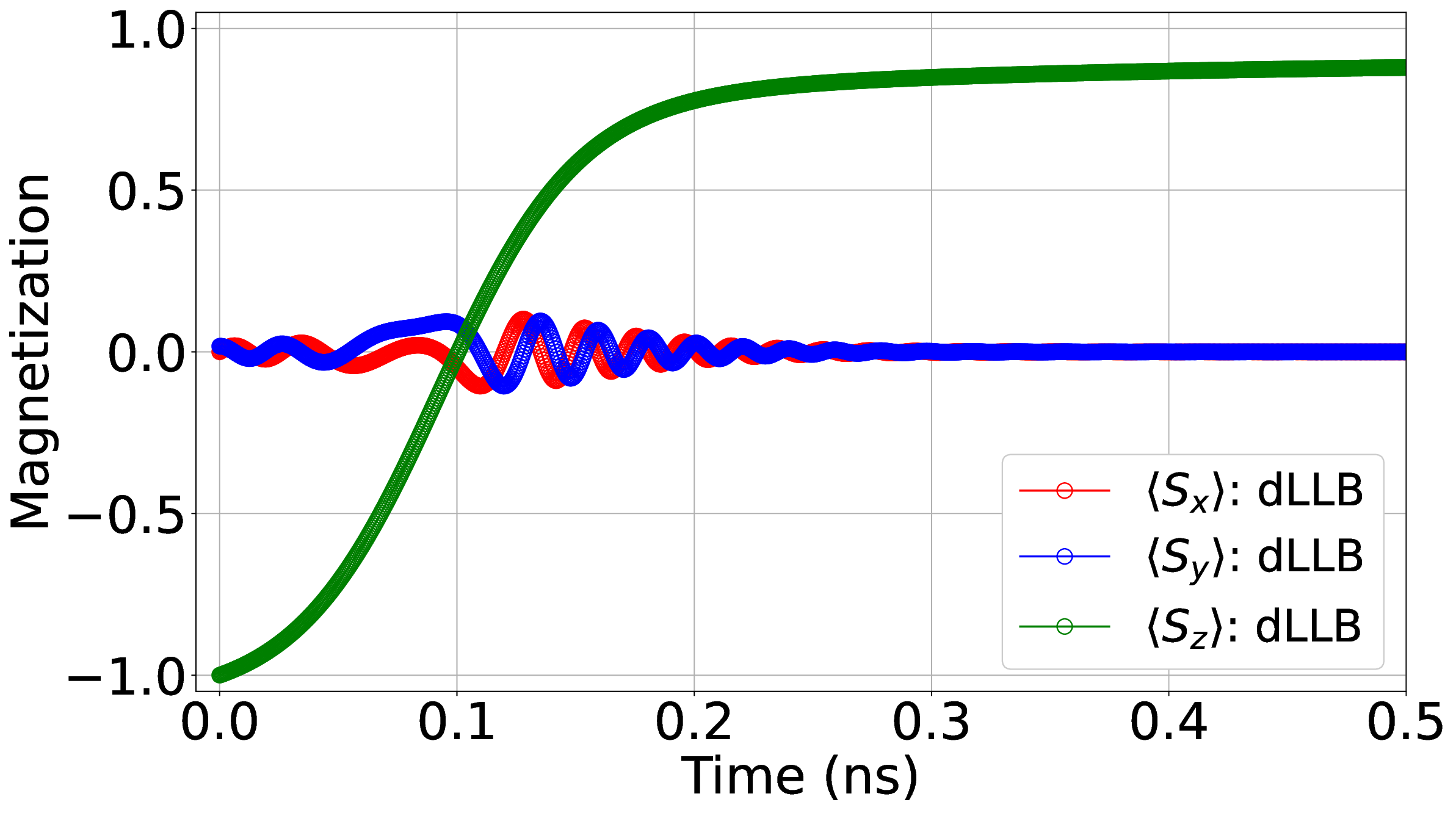}\\
    \end{tabular}
    \caption{Switching of an MTJ induced by a (damping-like) spin-transfer torque, simulated via the dLLBS approach for increasing operating temperatures, in the presence of a magnetic anisotropy axis in the free-layer. (see text).}
    \label{fig:dLLBS_switch}
\end{figure}

Second, we consider the spin-transfer torque switching.
In the free layer, a magnetic anisotropy field along the direction ${\bm n}=\hat{\bm z}$ is included, in addition to a pure damping-like spin-transfer polarization also along the $\hat{\bm z}$-axis (i.e. $a_T=0.1\unit{\tesla}$, ${\bm p}=\hat{\bm z}$, $\alpha=0.005$ and ${\bm B}_K=B_K\hat{\bm z}$ with $B_K=1.5\unit{\tesla}$).
The figure~\ref{fig:dLLBS_switch} displays the resulting dynamics of a single dLLBS simulation, when the temperature is increasing.
In each simulation, the magnetization is initially aligned along the $-{\bm z}$-axis.
This result is consistent with both previous experimental and theoretical studies, which report sub-nanosecond switching timescales due to the strong internal fields.
The upper graph of the figure~\ref{fig:dLLBS_switch} displays the same physical content to the Figure 2 b) of Reference~\cite{meoSpintransferSpinorbitTorques2023}.
The central point here is that the lower graphs of figure~\ref{fig:dLLBS_switch} are produced several order of magnitude faster than any equivalent sLLG simulations.
Additionally and around $0.1\unit{\nano\second}$, there has been an almost complete disappearance of the average magnetization due to the competition between the transverse and longitudinal (thermally) torques. This scenario does not arise in deterministic magnetization dynamics governed by norm-preserving equations.

The stochastic nature of spin-transfer torque-induced magnetization dynamics have been considered in computing frameworks such as probabilistic computing and Bayesian logic \cite{jiaSPINBISSpintronicsBasedBayesian2020,camsariPbitsProbabilisticSpin2019}. In this regard, the dLLBS model offers a notable capability. Beyond providing efficient estimates of the average magnetic moment, the model explicitly tracks the temporal evolution of its variance through the second moment, $\Sigma(t)$ (Eq.~\eqref{dLLB2}). This allows for a dynamic description of the probability distribution of the magnetization, capturing statistical trends in the system with lower computational overhead compared to the sLLG approach.

From the perspective of spintronic applications, this capability is relevant to probabilistic computing. In such systems, functionality depends not only on the mean state of the magnetic ensemble but also on fluctuations and correlations that shape the probability landscape. The dLLBS framework, by capturing both the average trajectory and its variance, provides a means to study noise-driven phenomena, stochastic resonance, and distribution-based logic operations, which are central to Bayesian inference \cite{jiaSPINBISSpintronicsBasedBayesian2020,shimStochasticSpinOrbitTorque2017}. This approach enables the connection between device-level magnetization dynamics and system-level architectures in probabilistic spintronics, allowing for the exploration of design spaces where randomness plays a functional role.

\section{Conclusion}\label{sec:conclusion}

In this work, we presented the derivation of the dLLBS equations for spintronic devices, which systematically incorporate thermal fluctuations and spin torque effects. We analyzed their impact on switching dynamics in benchmark systems, demonstrating consistency with established models while capturing additional stochastic features. 
The model’s ability to account for both mean behavior and statistical variance highlights its advantages over traditional approaches like sLLG or LLB. 
Furthermore, we showcased the predictive power of the dLLBS framework for switching times in technologically relevant systems, such as MTJs. 
\textcolor{black}{For instance, in MTJs with perpendicular magnetic anisotropy, increasing temperature reduces both the saturation magnetization ($M_s$) and the effective anisotropy ($K_u$ for an uniaxial magnet). 
This reduction lowers the switching threshold (current/voltage) by decreasing the energy barrier, which scales roughly with the ratio of anisotropy energy to thermal energy ($\approx K_uV/k_BT$). 
Given that the dLLB model accurately reproduces thermal variations in saturation magnetization, as mentioned in Ref.~\cite{fattouhiUltrafastDynamicsMoments2025}, we hypothesize that it can similarly capture thermal effects on magnetic anisotropy -- and thus on the energy barrier. 
However, this requires dedicated investigation.}
These results underscore the dLLBS framework’s potential for optimizing spintronic devices and exploring noise-driven functionalities in probabilistic computing.

\appendix

\section{Derivation of the noise-averaged sLLG equations}
\label{appendix:derivation}
\textcolor{black}{
The derivation of the dLLBS equations requires a systematic noise-averaging of the sLLG equation.
This can be achieved by first rewriting the sLLG equation in the standard form of an ordinary differential equation—specifically, by isolating the first-order time derivative of ${\bm s}$ on the left-hand side. 
This algebraic manipulation follows a conventional approach, involving the expansion of the cross-product term ${\bm s}\times$ across both sides of Eq.~\eqref{sLLG} and by regrouping terms.
This recasts the equation in the Landau-Lifshitz damping form:
\begin{align}
    \begin{aligned}
        (1+\alpha^2)\frac{d{\bm s}}{dt}&=\bm{s}\times(\bm{\omega}(\bm{s})+\bm{h}(t))+\bm{T}_{\mathsf{ST}}(\bm{s})\\
        &+\alpha\bm{s}\times(\bm{s}\times(\bm{\omega}(\bm{s})+\bm{h}(t))+\bm{T}_{\mathsf{ST}}(\bm{s}))
    \end{aligned}
    \label{appendix:sLL}
\end{align}  
Now we observe that the relevant torque is of the form $\bm{s}\times(\bm{\omega}(\bm{s})+\bm{h}(t))+\bm{T}_{\mathsf{ST}}(\bm{s})$.
In this expression, $\bm{\omega}(\bm{s})\equiv\gamma\bm{B}_{DC}+\gamma B_K(\bm{s}\cdot\bm{n})\bm{n}$ is the internal field of the free layer and contains both DC (Zeeman-like) and anisotropy (easy axis $\bm{n}$ and amplitude $B_K$) fields. 
Since the spin-transfer torque $\bm{T}_{\mathsf{ST}}(\bm{s})$ in Eq.~\eqref{eq:STTtorque} represents an external torque applied to the free layer, comprising both a \textit{field-like} and a \textit{damping-like} component relative to a fixed polarization $\bm{p}$, we express the total effective field in Eq.~\eqref{appendix:sLL} as:
\begin{widetext}
    \vspace{-\baselineskip} 
\begin{align}
    \begin{aligned}
        \bm{\omega}(\bm{s})+\bm{h}(t)+a_T\bm{s}\times\bm{p}+b_T\bm{p}&=\gamma\bm{B}_{DC}+b_T\bm{p}+\bm{h}(t)+(\gamma B_K\bm{n}\otimes\bm{n}-a_T[\bm{p}]_\times)\bm{s}\\
    &=\bm{u}+\bm{h}(t)+\mathfrak{M}\bm{s}
    \end{aligned}
\end{align}
\end{widetext}
where we define $\bm{u}\equiv\gamma\bm{B}_{DC}+b_T\bm{p}$ as a pulsation vector independent of $\bm{s}$ and $\mathfrak{M}\equiv\gamma B_K\bm{n}\otimes\bm{n}-a_T[\bm{p}]_\times$ a matrix containing a symmetric and antisymmetric parts, also independent of $\bm{s}$ that couples linearly with it. 
In these expressions $\bm{h}(t)$ is the only explicit stochastic (thermal) field.}

\textcolor{black}{ 
To simplify the noise-averaging process, we reformulate Eq.~\eqref{appendix:sLL} as follows:
\begin{align}
    (1+\alpha^2)\frac{d{\bm s}}{dt}&=\bm{A}(\bm{s})+\mathfrak{E}(\bm{s})\bm{h}(t)
\end{align}
where $\bm{A}$ is an explicit vector of $\bm{s}$ and $\mathfrak{E}$ is an explicit matrix of $\bm{s}$ (also known as the vielbein matrix in the tetrad formalism). These quantities have the following forms:
\begin{align}
    \bm{A}(\bm{s})&\equiv\bm{s}\times(\bm{u}+\mathfrak{M}\bm{s})+\alpha\bm{s}\times\left(\bm{s}\times(\bm{u}+\mathfrak{M}\bm{s}\right)\label{appendix:eqA}\\
    \mathfrak{E}(\bm s)&\equiv[\bm{s}]_\times+\alpha[\bm{s}]_\times[\bm{s}]_\times\label{appendix:eqE}
\end{align}
By following Ref.\cite{tranchidaHierarchiesLandauLifshitzBlochEquations2018}, the noise-average is now straightforward, and we have these equations for the first and second moment of the magnetization:
\begin{align}
    \label{appendix:s1}
    (1+\alpha^2)\frac{d\langle\bm{s}\rangle}{dt}&=\langle\bm{A}(\bm{s})\rangle-\frac{2D}{1+\alpha^2}\langle\bm{s}\rangle\\
    \label{appendix:s2}
    (1+\alpha^2)\frac{d\langle\bm{s}\otimes\bm{s}\rangle}{dt}&=\langle\bm{A}(\bm{s})\otimes\bm{s}\rangle+\langle\bm{s}\otimes\bm{A}(\bm{s})\rangle\nonumber\\
    &-\frac{2D}{1+\alpha^2}\left(3\langle\bm{s}\otimes\bm{s}\rangle-\textrm{Tr}(\langle\bm{s}\otimes\bm{s}\rangle)\mathbbold{1}\right)
\end{align}
where the term proportional to $\alpha$ in Eq.~\eqref{appendix:eqE} can be safely neglected (see Ref.\cite{bertottiNonlinearMagnetizationDynamics2009}).
These expressions form an open hierarchy of coupled differential equations, which must be closed for consistent solution. 
This is achieved via the Gaussian Closure Approximation, which assumes that the third and fourth cumulants of $\bm{s}$ are zero.
In equation, this reads:
\begin{align}
    \begin{aligned}
        0=&\langle\!\langle s_is_js_l\rangle\!\rangle \\
        =&\langle s_is_js_l\rangle-\langle s_is_j\rangle\langle s_l\rangle-\langle s_is_l\rangle\langle s_j\rangle\\
        &-\langle s_js_l\rangle\langle s_i\rangle+2\langle s_i\rangle\langle s_j\rangle\langle s_l\rangle
    \end{aligned}
\label{appendix:Cumulants3}
\end{align}
for any combination of the vector indices of $\bm{s}$ for the third-order cumulant and 
\begin{align}
    \begin{aligned}
        0&=\langle\!\langle s_is_js_ls_m\rangle\!\rangle \\
        &=\langle s_is_js_ls_m\rangle-6\langle s_i\rangle\langle s_j\rangle\langle s_l\rangle\langle s_m\rangle-\langle s_is_j\rangle\langle s_ls_m\rangle\\
        &-\langle s_is_l\rangle\langle s_js_m\rangle-\langle s_is_m\rangle\langle s_js_l\rangle-\langle s_i\rangle\langle s_js_ls_m\rangle\\
        &-\langle s_j\rangle\langle s_is_ls_m\rangle-\langle s_l\rangle\langle s_is_js_m\rangle-\langle s_m\rangle\langle s_is_js_l\rangle\\
        &+2\left\{\langle s_i\rangle\langle s_j\rangle\langle s_ls_m\rangle+\langle s_i\rangle\langle s_l\rangle\langle s_js_m\rangle+\langle s_i\rangle\langle s_m\rangle\langle s_js_l\rangle\right.\\
        &+\left.\langle s_j\rangle\langle s_l\rangle\langle s_is_m\rangle+\langle s_j\rangle\langle s_m\rangle\langle s_is_l\rangle+\langle s_l\rangle\langle s_m\rangle\langle s_is_j\rangle\right\}
    \end{aligned}
    \label{appendix:Cumulants4}
\end{align}
for any combination of the vector indices for the fourth-order cumulant.
Thus, the third- and fourth-order moments of $\bm{s}$ must be substituted from Eqs.\eqref{appendix:Cumulants3} and \eqref{appendix:Cumulants4} into Eqs.\eqref{appendix:s1} and \eqref{appendix:s2}. After further algebraic simplifications, this yields Eqs.~\eqref{dLLB1} and \eqref{dLLB2}.
}

\section*{Data Availability Statement}\label{sec:DAL}
The data that support the findings of this study are available from the corresponding author, [PT], upon reasonable request.

\section*{Acknowledgments}\label{sec:acknowledgments}
This work is partially supported by the France 2030 government investment plan managed by the French National Research Agency under grant reference PEPR SPIN – [SPINTHEORY] ANR-22-EXSP-0009, and CEA Exploratory Program, Bottom-Up 2023 ref.59 [THERMOSPIN].

%
    
\end{document}